\documentclass{fairmeta}

\usepackage{amsmath}
\usepackage{amssymb}
\usepackage{enumitem}
\usepackage{subcaption}
\usepackage{tikz}
\usetikzlibrary{positioning, arrows.meta, shadows, backgrounds, calc, shapes.geometric, fit}
\usepackage{balance}

\usetikzlibrary{arrows.meta,positioning}
\newcommand{\prism}{PRISM}
\newcommand{\R}{\mathbb{R}}
\newcommand{\sg}{\operatorname{sg}}

\title{PRISM: Personalized Item Representations in User History}

\author[1]{Renzhi Wu}
\author[1]{Zikun Cui}
\author[1]{Junjie Yang}
\author[1]{Li Sheng}
\author[1]{Yongjun Xie}
\author[2]{Jerry Fu}
\author[1]{Nipun Mathur}
\author[1]{Sri Reddy}
\author[1]{Hong Li}
\affiliation[1]{Meta MRS}
\affiliation[2]{Facebook Discovery}

\abstract{
Sequential recommenders typically construct history inputs from item
identifiers or content features using an embedding map shared across users.
The same item therefore yields the same initial embedding representation,
even when users have different relationships to the item. We propose
\emph{personalized item representations in user history}, where each item's
embedding expresses its relationship to the user.
Personalization becomes a property of the history inputs themselves, supplying richer evidence
for later sequence modeling. We instantiate this idea with \prism{}, which
reuses upstream ranking computation to generate personalized item embeddings.
An autoencoder jointly compresses all layer states from a user tower, a user--item
tower, and same-request candidate mixing into one embedding per scored pair
for downstream replay.
Offline evaluation on Surface~1, a major recommendation
surface at billion scale, reports an average downstream normalized entropy
(NE) change of
$-0.15\%$, compared with $-0.07\%$ for VISTA at similar history lengths.
We further identify an empirical sequence-length scaling law: over history
lengths of 26--240 events, NE gain follows a fitted power law with exponent $0.468$, increasing
from $0.06\%$ to $0.17\%$. Therefore, \emph{personalized item representations in user history} provide
a new scaling dimension for improving recommendation quality.
}

\date{\today}
\correspondence{Renzhi Wu at \email{renzhiwu@meta.com}}
\metadata[Keywords]{Sequential Recommendation, Sequence Modeling}

\begin{document}
\maketitle

\section{Introduction}

Sequence modeling is central to personalized recommendation. Existing models,
including SASRec~\citep{sasrec}, HSTU~\citep{hstu}, CMSL~\citep{cmsl}, and
OneTrans~\citep{onetrans}, share a common starting point: each item in user
history is mapped to an embedding before sequence processing, using item-side
features such as item IDs, creator IDs, and content descriptors. The embedding
map is shared across users, so the same item receives
the same initial embedding in different users' histories. After this step,
sequence encoders and target-aware interaction learn user-dependent states
from these inputs~\citep{din,dien}.

\begin{figure*}[t]
\centering
\begingroup
\definecolor{historyink}{HTML}{243746}
\definecolor{historymuted}{HTML}{62717C}
\definecolor{historypanel}{HTML}{F4F6F8}
\definecolor{historyA}{HTML}{0072B2}
\definecolor{historyB}{HTML}{D58A00}
\definecolor{historyC}{HTML}{009E73}
\definecolor{historyD}{HTML}{A75A8C}
\newcommand{\historytoken}[7]{%
  \node[token] (#1) at (#2,#3) {};
  \if\relax\detokenize{#7}\relax
    \node[font=\footnotesize\bfseries,text=#5] at (#2,#3+0.23) {#4};
  \else
    \node[font=\footnotesize\bfseries,text=#5] at (#2-0.32,#3+0.23) {#4};
    \node[font=\scriptsize,text=historymuted] at (#2+0.25,#3+0.23) {#7};
  \fi
  \foreach \h [count=\j from 0] in {#6} {
    \fill[#5!78] (#2-0.37+0.16*\j,#3-0.24)
      rectangle ++(0.105,\h);
  }
}
\begin{tikzpicture}[
  font=\footnotesize,text=historyink,
  token/.style={draw=historyink!24,fill=white,rounded corners=3pt,
    minimum width=13mm,minimum height=9mm,inner sep=0pt},
  flow/.style={-{Latex[length=1.65mm]},draw=historyink!45,line width=0.55pt}]
 \fill[historypanel,rounded corners=5pt] (0,0) rectangle (8.25,3.2);
 \fill[historypanel,rounded corners=5pt] (8.65,0) rectangle (16.9,3.2);
 \node[anchor=west,font=\footnotesize\bfseries] at (0.25,2.94)
   {(a) Shared item history};
 \node[anchor=west,font=\footnotesize\bfseries] at (8.90,2.94)
   {(b) PRISM: personalized item history};
 \node[anchor=west,text=historymuted] at (0.25,2.58)
   {What is the item?};
 \node[anchor=west,text=historymuted] at (8.90,2.58)
   {How does this user relate to the item?};

 \foreach \x/\y/\u/\role in
   {0.35/1.92/X/Casual viewer,0.35/0.87/Y/Food creator,
    9.00/1.92/X/Casual viewer,9.00/0.87/Y/Food creator} {
   \node[anchor=west,font=\footnotesize\bfseries] at (\x,\y+0.13) {User $\u$};
   \node[anchor=west,font=\scriptsize,text=historymuted] at (\x,\y-0.16) {\role};
 }

 \historytoken{xa}{2.65}{1.92}{$A$}{historyA}{0.17,0.30,0.23,0.11,0.27}{$e_A$}
 \historytoken{xb}{4.55}{1.92}{$B$}{historyB}{0.27,0.12,0.20,0.31,0.16}{$e_B$}
 \historytoken{xd}{6.45}{1.92}{$D$}{historyD}{0.13,0.25,0.31,0.18,0.22}{$e_D$}
 \historytoken{yc}{2.65}{0.87}{$C$}{historyC}{0.30,0.19,0.12,0.26,0.15}{$e_C$}
 \historytoken{ya}{4.55}{0.87}{$A$}{historyA}{0.17,0.30,0.23,0.11,0.27}{$e_A$}
 \historytoken{yd}{6.45}{0.87}{$D$}{historyD}{0.13,0.25,0.31,0.18,0.22}{$e_D$}

 \historytoken{pxa}{11.30}{1.92}{$A$}{historyA}{0.30,0.11,0.24,0.17,0.28}{}
 \historytoken{pxb}{13.20}{1.92}{$B$}{historyB}{0.15,0.29,0.12,0.22,0.30}{}
 \historytoken{pxd}{15.10}{1.92}{$D$}{historyD}{0.26,0.14,0.30,0.11,0.19}{}
 \historytoken{pyc}{11.30}{0.87}{$C$}{historyC}{0.12,0.26,0.18,0.31,0.21}{}
 \historytoken{pya}{13.20}{0.87}{$A$}{historyA}{0.11,0.24,0.15,0.30,0.18}{}
 \historytoken{pyd}{15.10}{0.87}{$D$}{historyD}{0.16,0.30,0.12,0.23,0.28}{}

 \foreach \a/\b in {xa/xb,xb/xd,yc/ya,ya/yd,pxa/pxb,pxb/pxd,pyc/pya,pya/pyd}
   \draw[flow] (\a.east)--(\b.west);
 \foreach \n in {xa,xb,xd,yc,ya,yd,pxa,pxb,pxd,pyc,pya,pyd}
   \draw[historyink!18] (\n.center) ++(-0.42,-0.24)--++(0.84,0);

 \node[font=\scriptsize\bfseries] at (4.125,0.20)
   {Same cooking video $A$: one shared item vector};
 \node[font=\scriptsize\bfseries] at (12.775,0.20)
   {Same cooking video $A$: different user perspectives};
\end{tikzpicture}
\endgroup
\caption{From shared item history to personalized item history. A cooking video
($A$) has the same initial item vector for a casual viewer and a food creator.
PRISM represents historical items in relation to the user, supplying personalized
evidence to a later sequence model. Hue identifies the item; vector glyphs
schematically illustrate shared versus personalized inputs.}
\label{fig:representations}
\end{figure*}
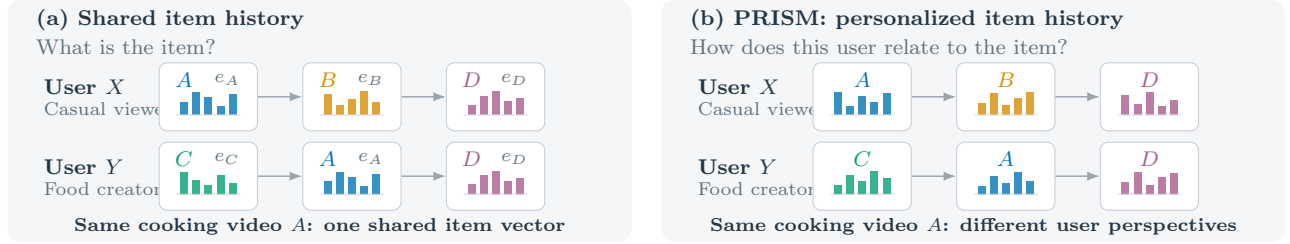

\noindent\textbf{Item embeddings: from shared to personalized.}
The above mentioned standard practice limits personalization at the input stage.
For example, a cooking video may be occasional entertainment for a casual
viewer but a recurring source of ideas for a food creator. A shared initial
embedding does not express these different relationships. Motivated by this
limitation, we propose \emph{personalized item representations in user history}.
As illustrated in Figure~\ref{fig:representations}, each history entry represents
how an item relates to the user. In the cooking video example,
personalized embeddings can distinguish entertainment from professional interest.
Furthermore, even for the same user, the same item can receive different
embeddings across repeated events as interests and context evolve. This
personalizes the initial history inputs for each user and each moment, giving
sequence models richer evidence for learning user preferences.

However, our proposal introduces a scalability challenge. A naive approach would
train and run a separate user--item model to generate rich embeddings across
a large volume of historical events. We observe that recommendation systems
often already use an upstream ranker to provide learned user embedding
features and teacher labels for downstream models. Such a ranker reads user
history and combines user, item, and context features to predict engagement;
its hidden states provide learned evidence about each user--item encounter.
Therefore, we retain this computation in a compact embedding per scored pair and chronologically replay these embeddings to construct a personalized
history for later sequence modeling, amortizing generation computation.

\noindent\textbf{The three complementary views through PRISM.}
We introduce \prism{}, an upstream ranker designed to generate personalized
item embeddings for downstream sequence modeling. Its representation learning
captures three complementary views. \emph{User state} describes the user's
interests at the moment of viewing the item. \emph{User--item interaction} describes
how the item relates to that user state. \emph{Session interaction} describes the
item's relationships with other candidates in the same session, providing
evidence about the user's immediate context. Together, these views express
the user--item relationship comprehensively at the moment of the historical encounter.

\begin{figure*}[t]
\centering
\begingroup
\definecolor{methodink}{HTML}{243746}
\definecolor{methodmuted}{HTML}{62717C}
\definecolor{methodpanel}{HTML}{F4F6F8}
\definecolor{methoduser}{HTML}{0072B2}
\definecolor{methodpair}{HTML}{D58A00}
\definecolor{methodsession}{HTML}{A75A8C}
\definecolor{methodcode}{HTML}{009E73}
\begin{tikzpicture}[
  font=\footnotesize,text=methodink,
  tower/.style={draw=methodink!24,fill=white,rounded corners=3pt,
    minimum width=41mm,minimum height=13mm,inner sep=0pt},
  states/.style={rounded corners=2pt,minimum width=34mm,
    minimum height=4.6mm,inner sep=1pt,align=center},
  box/.style={draw=methodink!24,fill=white,rounded corners=3pt,
    align=center,inner sep=4pt,minimum height=8mm},
  flow/.style={-{Latex[length=1.65mm]},draw=methodink!45,line width=0.55pt},
  tap/.style={-{Latex[length=1.65mm]},line width=0.65pt},
  training/.style={-{Latex[length=1.65mm]},draw=methodink!45,
    dashed,line width=0.55pt}]

 \fill[methodpanel,rounded corners=5pt] (0,1.70) rectangle (16.9,4.72);
 \fill[methodpanel,rounded corners=5pt] (0,-0.42) rectangle (16.9,1.58);

 \node[text=methodmuted] (uin) at (2.45,4.42) {User features + history};
 \node[text=methodmuted] (pin) at (8.45,4.42) {Candidate features + context};
 \node[text=methodmuted] (sin) at (14.45,4.42) {Same-request candidate set};
 \node[tower] (user) at (2.45,3.50) {};
 \node[tower] (pair) at (8.45,3.50) {};
 \node[tower] (session) at (14.45,3.50) {};
 \node[font=\footnotesize\bfseries,text=methoduser] at (2.45,3.93) {User tower};
 \node[font=\footnotesize\bfseries,text=methodpair] at (8.45,3.93) {User--item tower};
 \node[font=\footnotesize\bfseries,text=methodsession] at (14.45,3.93) {Session mixing};
 \node[font=\scriptsize,text=methodmuted] at (2.45,3.59) {User state at the event};
 \node[font=\scriptsize,text=methodmuted] at (8.45,3.59) {User--item relationship};
 \node[font=\scriptsize,text=methodmuted] at (14.45,3.59) {Context from same-request items};
 \node[states,fill=methoduser!11,text=methoduser] (umid) at (2.45,3.18)
   {$(H_r^{U,\ell})_{1\leq\ell\leq L_U}$};
 \node[states,fill=methodpair!12,text=methodpair] (pmid) at (8.45,3.18)
   {$(H_e^{P,\ell})_{1\leq\ell\leq L_P}$};
 \node[states,fill=methodsession!12,text=methodsession] (smid) at (14.45,3.18)
   {$(H_e^{S,\ell},s_r^\ell)_{1\leq\ell\leq L_S}$};
 \draw[flow] (uin.south)--(user.north);
 \draw[flow] (pin.south)--(pair.north);
 \draw[flow] (sin.south)--(session.north);
 \draw[flow] (user.east)--node[above,font=\scriptsize,align=center]
   {Final user\\state}(pair.west);
 \draw[flow] (pair.east)--node[above,font=\scriptsize,align=center]
   {Detached\\final pairs}(session.west);

 \node[box,minimum width=161mm] (gather) at (8.45,2.22)
   {\textbf{All layers} $\longrightarrow$ stop-gradient
    $\longrightarrow$ normalize each part $\longrightarrow$ concatenate $v_e$};
 \draw[tap,draw=methoduser] (umid.south)--(umid.south |- gather.north);
 \draw[tap,draw=methodpair] (pmid.south)--(pmid.south |- gather.north);
 \draw[tap,draw=methodsession] (smid.south)--(smid.south |- gather.north);

 \node[box,text width=31mm] (encoder) at (2.10,1.05)
   {\textbf{Encoder}\\compress $v_e$};
 \node[box,text width=15mm] (code) at (6.40,1.05)
   {\textbf{Event code}\\\textcolor{methodcode}{$p_e$}};
 \node[box,text width=30mm] (export) at (10.70,1.05)
   {\textbf{Event store}\\$p_e$ + IDs/time/version};
 \node[box,text width=29mm] (replay) at (14.85,1.05)
   {\textbf{Later request}\\history reader};
 \draw[flow] (gather.south -| encoder.north)--(encoder.north);
 \draw[flow] (encoder.east)--(code.west);
 \draw[flow] (code.east)--node[above,font=\scriptsize]{export}(export.west);
 \draw[flow] (export.east)--node[above,font=\scriptsize]{replay}(replay.west);

 \node[box,dashed,text width=33mm,
   font=\scriptsize,minimum height=6mm] (decoder) at (6.40,0.12)
   {\textbf{Decoder: training only}\\reconstruct $v_e$; $\mathcal L_{\rm AE}$};
 \draw[training] (code.south)--(decoder.north);
 \node[anchor=west,font=\scriptsize,text=methodmuted,align=left] at (9.10,0.12)
   {Colored taps retain all layer states from all three sources.\\
    Reconstruction updates only the encoder and decoder.};
\end{tikzpicture}
\endgroup
\caption{Complete PRISM architecture. User states seed the pair workspace;
detached pair states feed session mixing. Colored taps gather every
layer, including contextual and pooled session states. An autoencoder
compresses them for event export and downstream replay. Its decoder learns
reconstruction during training.}
\label{fig:method}
\end{figure*}
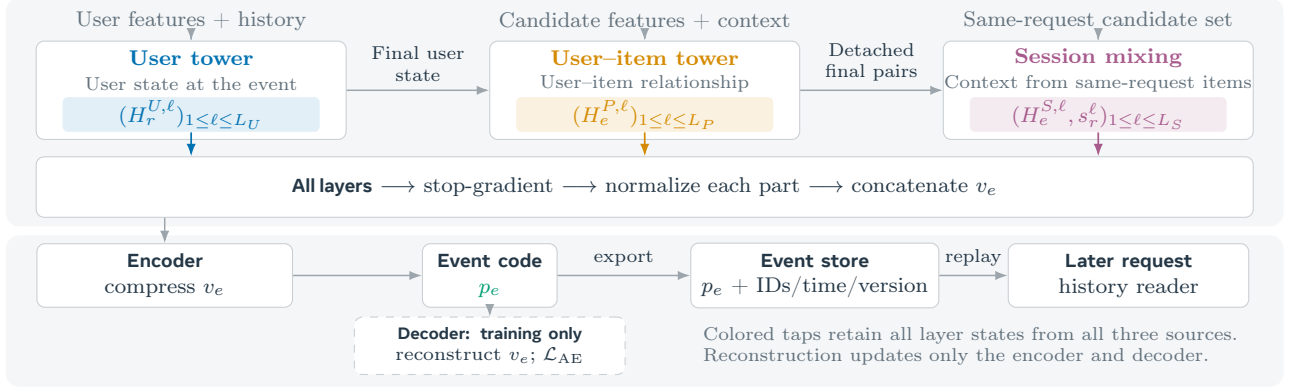

Figure~\ref{fig:method} maps these views to a dedicated \emph{user tower}, a
\emph{user--item tower}, and \emph{same-request candidate mixing}. The user
tower reads prior history and other user features; its states also inform
pair modeling. Candidate mixing contextualizes each item through the other
items in its request. A detached autoencoder jointly compresses their
\emph{hidden states from every layer}, including candidate-specific and pooled
session states, into one personalized embedding. Joint compression allocates
one code budget across sources and depths; downstream models learn how to
read the retained evidence for their own prediction objectives.

We evaluate PRISM history as an additional input to the ranker for Surface~1,
a major recommendation surface at billion scale. PRISM achieves an average
normalized entropy (NE; lower is better) change of $-0.15\%$,
compared with $-0.07\%$ for VISTA~\citep{vista} using similar
history lengths. We further identify an empirical \emph{sequence-length
scaling law} (Figure~\ref{fig:scaling}): NE gain follows a fitted power law
$g(L)\propto L^{0.468}$. The sublinear exponent
captures improving recommendation quality per additional event at log scale.
Therefore, \emph{personalized item representations in user history} provide
a \emph{new scaling dimension} for improving recommendation quality.

We make four contributions:
\begin{itemize}[leftmargin=*]
 \item \textbf{Personalized item representations in user history.} We introduce
 a new history input design for sequential recommendation in which each item's representation is personalized to the user, allowing the same item to differ across users and time.
 \item \textbf{Embedding generation through upstream reuse.} We instantiate this
 idea with an efficient design, which exports compact embeddings from scored pairs and
 replays them as history, amortizing upstream ranking computation across future reads.
 \item \textbf{PRISM's three modeling components.} We propose the PRISM upstream user model that captures the three complementary views through a
 user tower, a user--item tower, and same-request candidate mixing.
 A detached autoencoder then jointly compresses states from every layer into one personalized embedding.
 \item \textbf{An empirical scaling law for personalized history.} We identify
 a scaling law relationship between personalized history length and downstream ranking quality, providing a new scaling dimension for improving
 recommendation systems.
\end{itemize}

\section{Related Work}

PRISM focuses on personalizing history inputs before downstream sequence
processing. Sequence encoders~\citep{sasrec,bert4rec,s3rec,hstu} and interest
models~\citep{din,dien} learn user-dependent states from behavioral inputs;
CMSL~\citep{cmsl} constructs multiple latent history contexts. PRISM supplies
these models with an item representation specific to each historical encounter
with the user. Reuse across recommendation stages connects our work to learned
user representations~\citep{pancha2022pinnerformer}, multistage sequence
modeling~\citep{llatte}, and graph representation learning~\citep{rankgraph,rankgraph_2}.
The closest system comparisons are VISTA~\citep{vista}, which caches user-history
summary tokens for downstream target attention, and OneTrans~\citep{onetrans},
which reuses intermediate states through cross-request KV caching. PRISM
stores a compact code for each scored user--item pair, allowing later readers
to select individual events with the user and context evidence retained at
each encounter. Session recommenders such as GRU4Rec and SR-GNN model observed
interaction sessions~\citep{gru4rec,srgnn}; PRISM's session component contextualizes
candidates in the same ranking request, drawing on set-context modeling
ideas~\citep{zaheer2017deepsets,lee2018settransformer}. Feature-interaction
architectures~\citep{dlrm,dcn,dcnv2,wukong} and latent-array
processing~\citep{perceiver} provide precedent for the rich computation that
PRISM retains. Distillation transfers prediction knowledge~\citep{distillation},
while FitNets and Hypercolumns exploit intermediate representations as hints
or descriptors~\citep{romero2014fitnets,hariharan2014hypercolumns}; autoencoders
support dimensionality reduction~\citep{hinton2006autoencoder} and collaborative
filtering~\citep{liang2018multivae}. PRISM jointly compresses states from every
layer of its user, user--item, and same-request components into personalized
history tokens for later supervised reading. Complementing recommendation
scaling studies~\citep{wukong,llatte}, our experiments examine the gain from
retaining longer histories of these personalized event codes.

\section{Method}

Figure~\ref{fig:method} shows the complete implementation.
Let $e=(u,i,\tau,r)$ identify user, item, scoring time, and request.
Let $x_u^{<\tau}$ be user history before $\tau$, $f_u^\tau$ other user
features, $x_i$ candidate features, and $c_r$ shared request context.
The generation interface is
$p_e=F_\theta(x_u^{<\tau},f_u^\tau,x_i,c_r)\in\R^{d_p}$: a fixed-width code computed
from information available at the historical event, where $\theta$ denotes
the generator's parameters and $d_p$ the exported embedding width. Its user and context
inputs remain attached to that event. A later reader uses its new candidate
and current context to select saved codes.

\subsection{User and User--Item Towers}

Let $L_U$ and $L_P$ be the user and pair tower depths; $\ell$ indexes layers,
and superscripts $U$ and $P$ identify the two components.
The user tower reads the history together with candidate-independent sparse
feature embeddings, dense embedding features, and continuous features.
It produces $H_r^{U,1},\ldots,H_r^{U,L_U}$ from these user inputs and shared
request context, where $H_r^{U,\ell}$ is the hidden-state tensor at layer $\ell$.
Its final
state is shared across the request's candidates and incorporated into the
initial user--item workspace. Thus user evidence also participates directly
in pair-state learning.

The pair tower produces hidden states $H_e^{P,1},\ldots,H_e^{P,L_P}$ for
event $e$. In the model
backbone, blocks perform self-attention, candidate-conditioned reads of history
and feature sources, and higher-order interaction. Condensates from preceding
blocks remain available to later blocks. Final pair states feed multitask
prediction heads, trained with
$\mathcal L_P=\sum_e\sum_m w_{e,m}\,
\ell_m\big(y_{e,m},h_m(H_e^{P,L_P})\big)$.
The sum is over training events $e$; $m$ indexes item prediction tasks,
$y_{e,m}$ is the observed target,
$h_m$ the pair prediction head, and $\ell_m$ its loss function.
The weight $w_{e,m}$ combines task weights and missing-label masks. Binary tasks
use cross entropy; other tasks retain their appropriate objectives. This
supervision learns the user and pair towers.

\subsection{Session Mixing Within a Request}

A session is the candidate set $\mathcal C_r=\{e_1,\ldots,e_{n_r}\}$ in one
ranking request, where $n_r$ is the number of valid candidates.
The session mixer $G_\psi$ has parameters $\psi$ and $L_S$ layers;
superscript $S$ denotes session states. The operator $\sg$ stops gradients.
This masked interaction module jointly processes detached final pair states:
\begin{equation}
 \big(\{H_{e_j}^{S,\ell}\}_{j=1}^{n_r}\big)_{\ell=1}^{L_S}
 =G_\psi\!\left(\{\sg(H_{e_j}^{P,L_P})\}_{j=1}^{n_r},c_r\right).
 \label{eq:session}
\end{equation}
A Transformer-style module is one realization. Each layer gives a
candidate-specific contextual state $H_e^{S,\ell}$ and a fixed-width request summary
$s_r^\ell=\operatorname{Pool}_{\mathrm{mask}}
\big(\{H_{e_j}^{S,\ell}\}_{j=1}^{n_r}\big)$.
Here $\operatorname{Pool}_{\mathrm{mask}}$ pools over valid candidates;
both interaction and pooling mask padded candidates. 

Item heads on $H_e^{S,L_S}$ predict the ordinary engagement labels. Aggregate
heads on $s_r^{L_S}$ predict request-level outcomes, such as total action
counts. Their weighted objective is
\begin{equation}
 \mathcal L_S={}\sum_{e,m}\widetilde w_{e,m}\ell_m(y_{e,m},g_m(H_e^{S,L_S}))
 +\sum_{r,a}\lambda_a\ell_a(Y_{r,a},g_a(s_r^{L_S})).
 \label{eq:sessionloss}
\end{equation}
Here $a$ indexes aggregate tasks, $Y_{r,a}$ is the request-level target,
and $g_m$ and $g_a$ are item and aggregate prediction heads.
The item loss $\ell_m$ is defined above; $\ell_a$ is the aggregate loss.
Weights $\widetilde w_{e,m}$ include item task weights and masks for missing
or ineligible labels,
and $\lambda_a$ weights aggregate tasks.

\subsection{Multi-Level Hidden-State Compression}

The exported representation combines \emph{all layers}. The
ordered collection is defined as
\begin{equation}
     \mathcal B_e={}(H_r^{U,\ell})_{\ell=1}^{L_U}
 \;\Vert\;(H_e^{P,\ell})_{\ell=1}^{L_P}
 \Vert\;(H_e^{S,\ell},s_r^\ell)_{\ell=1}^{L_S},
\end{equation}
where $\Vert$ concatenates ordered parts. User and pooled session states are
shared across candidates; pair and contextual session states are event-specific.
Flattened source widths sum to target dimension $D$. Each session term contains
only this candidate's state and its request summary, so $D$ is independent of
$n_r$. Code width $d_p$ sets the transfer budget independently.

The three sources capture complementary aspects of personalization. User
states describe the user's current interests at this event. Pair states
express how the user's state interacts with this item's features. Session
mixing places the candidate among other items in the request; candidate-specific
and pooled states preserve local relationships and shared context. Joint
compression combines these overlapping perspectives into one personalized
item embedding, retaining both common evidence and event-specific refinements.


For each part $h\in\mathcal B_e$, we apply normalization to a
detached copy,
$\rho(x)=\kappa x/\sqrt{\operatorname{mean}(x^2)+\epsilon}$, where $\kappa$
is a fixed scale, $\epsilon>0$ ensures numerical stability, and the mean is
over coordinates. We then
concatenate the vectors
to obtain $v_e=\operatorname{concat}_{h\in\mathcal B_e}
\rho\big(\sg(\operatorname{vec}(h))\big)$ where $\operatorname{vec}$ flattens a tensor;
Per-part normalization controls depth-dependent scale differences;
Let $E_\omega:\R^D\to\R^{d_p}$ and $D_\omega:\R^{d_p}\to\R^D$ be the
encoder and decoder, with $\omega$ collecting their separate parameters.
With reconstruction weight $\lambda_{\rm AE}>0$, the multilevel compressor learns
\begin{equation}
 \begin{split}
 p_e&=\operatorname{softsign}(E_\omega(v_e)),\\
 \mathcal L_{\rm AE}&=\sum_e\frac{\lambda_{\rm AE}}{\dim(v_e)}
            \|v_e-D_\omega(p_e)\|_2^2.
 \end{split}
 \label{eq:ae}
\end{equation}
Here $\dim(v_e)=D$. The encoder and decoder use gated residual blocks with
RMS normalization. The encoder bounds coordinates with softsign,
$x/(1+|x|)$; the decoder reconstructs $D$ coordinates. We optimize
$\mathcal L_P+\mathcal L_S+\mathcal L_{\rm AE}$, with prediction losses learning
source states and reconstruction updating only the compressor.
The resulting $p_e\in\R^{d_p}$ fuses sources and depths, with different codes
possible for repeated items.




\subsection{Downstream Sequence Modeling}

Each stored event has an availability time $a_e$, when its code becomes
readable downstream, and a generator version $v$. At downstream time $t$,
select user events with $\tau<t$ and $a_e<t$ under a
specified retention policy, exclude the current request, and retain at most
$L$ history events.
Let $e_k$ be the $k$th retained event, with
scoring time $\tau_k$, generator version $v_k$, and eligible metadata $b_k$.
Let $d_r$ be the downstream reader width and
$A_{v_k}\in\R^{d_r\times d_p}$ the projection for version $v_k$.
The functions $E_{\rm time}$ and $E_{\rm event}$ encode elapsed time and
event metadata into $d_r$ coordinates. Construct a token
$z_k=A_{v_k}p_{e_k}+E_{\rm time}(t-\tau_k)+E_{\rm event}(b_k)$.
A sequence encoder and an attention
reader conditioned on the candidate consume these tokens alongside ordinary
features. Let $T$ stack the encoded tokens and $j$ be the current candidate.
The query network $Q_\phi$, with parameters $\phi$, combines current history
$x_u^{<t}$, user features $f_u^t$, candidate features $x_j$, and request context
$c_t$ to form $q_j=Q_\phi(x_u^{<t},f_u^t,x_j,c_t)$. Candidate attention returns
$o_j=\operatorname{Attn}(q_j,T,T)$.
Here $\operatorname{Attn}$ denotes attention with the query, key, and value
arguments shown, and $o_j$ is the candidate-conditioned history summary.
The predictor combines $o_j$ with ordinary features. Downstream supervision
updates the reader and projections; stored codes remain fixed. Missing history
uses an availability indicator and the ordinary path.





\section{Experiments}

\subsection{Setup and Metrics}
\noindent\textbf{Data.}
We use production data from Surface~1, a major recommendation surface at
billion scale. PRISM does online training for 14 consecutive days, exporting embeddings during training.
The downstream ranking models then train on three subsequent days of
production data, comprising approximately 30--50 billion examples, and
evaluate on the following day's held-out production data.

\noindent\textbf{Hardware.}
The PRISM model runs on 16 NVIDIA B200 nodes. Downstream models train
on 16 NVIDIA H100 GPUs using an internal industrial training platform,
mixed precision, and customized kernels for memory and throughput efficiency.

\noindent\textbf{Metrics.}
We use normalized entropy (NE)~\citep{he2014practical} as our metric. Smaller NE means better prediction performance.
Note an NE improvement greater than
$0.03\%$ is considered statistically significant.

\noindent\textbf{Settings.}
The PRISM arm adds the embedding history alongside the downstream ranker's
ordinary features; both baseline and PRISM arms train without warm start.
Exported codes have 128 dimensions, are INT8-quantized, and are accompanied
by timestamps. History retains the latest 400 events within
30 days; coverage is 98\%, with mean sequence length 235
and median 250.

\begin{table}[t]
\centering
\caption{Average NE changes (\%) versus the Surface~1 ranking
baseline, with similar history lengths for PRISM and VISTA. Lower is better.
VISTA's combined value is computed from its task-group means.}
\label{tab:results}
\setlength{\tabcolsep}{4pt}
\begin{tabular}{llc}
\toprule
Method & Task group & Average NE change (\%)\\
\midrule
VISTA & Positive (8) & $-0.05$\\
 & Negative (5) & $-0.09$\\
 & Combined (13) & $-0.07$\\
\midrule
PRISM & Positive (8) & $-0.17$\\
 & Negative (5) & $-0.11$\\
 & Combined (13) & $\mathbf{-0.15}$\\
\bottomrule
\end{tabular}
\par\medskip
\includegraphics[width=0.7\linewidth]{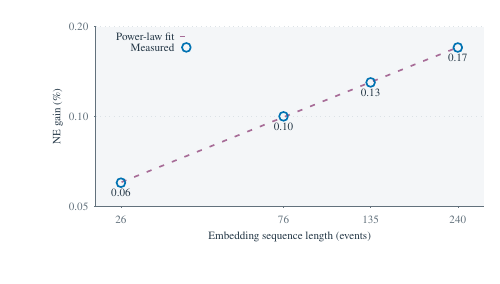}
\begingroup
\captionsetup*{type=figure}
\caption{Log--log sequence-length scaling. Circles show measurements;
the dashed line is the power-law fit $g(L)=0.0131L^{0.468}$, where $g$ is
NE gain (\%) and $L$ is history length.}
\label{fig:scaling}
\endgroup
\end{table}

\subsection{Downstream Ranking Results}

Table~\ref{tab:results} reports downstream average NE changes versus the
Surface~1 ranking baseline. Adding PRISM history yields changes of
$-0.17\%$ on positive engagement tasks and $-0.11\%$ on negative engagement tasks, with a combined
change of $-0.15\%$. Individual binary tasks also improve: comment
($-0.20\%$), like ($-0.22\%$), share ($-0.17\%$), and dislike ($-0.09\%$).
The gains across both task groups indicate that personalized history provides
useful predictive information beyond the ranker's ordinary features.

We also compare with VISTA~\citep{vista}, which caches user-history summary
tokens for downstream candidate attention. Both methods use the same
ranking baseline, data, ordinary features, prediction heads, and evaluation
protocol, with similar retained history lengths. VISTA achieves average NE
changes of $-0.05\%$ on positive engagement tasks and $-0.09\%$ on negative engagement tasks;
its combined change, weighted by task count, is $-0.07\%$. PRISM achieves
larger improvements in both groups, suggesting that preserving information
about individual historical encounters is valuable in this setting.

\subsection{Scaling Law with Sequence Length}

We next examine how downstream gains change as more personalized history is
retained. Figure~\ref{fig:scaling} shows NE improvements of $0.06\%$,
$0.10\%$, $0.13\%$, and $0.17\%$ for histories of 26, 76, 135, and 240
events, respectively. Fitting a power law $g(L)=aL^\beta$ to these measurements
by least squares in log space gives $g(L)\approx0.0131L^{0.468}$, where $L$
is the number of retained events and $g(L)$ is NE improvement in percent.
The fit appears as a straight line on the log--log axes. Its exponent
indicates diminishing gains per additional event: doubling history length
corresponds to approximately $1.38\times$ the NE improvement within the
observed range. This empirical relationship describes the tested
26--240-event regime and identifies retained personalized history as a
scaling dimension for recommendation. 

\section{Conclusion}
We introduced \emph{personalized item representations in user history},
enabling history inputs to express how each item relates to a particular
user at a particular encounter. PRISM implements this idea by reusing upstream
ranking computation: it jointly compresses states from every layer of its
user tower, user--item tower, and same-request candidate mixer into compact
event embeddings for downstream sequence modeling. Evaluation on Surface~1
reports an average NE change of $-0.15\%$, compared with $-0.07\%$ for VISTA
at similar history lengths. A sweep over 26--240 retained events further
reveals an empirical power-law trend with diminishing gains per additional
event. Together, these findings identify personalized history as a new
scaling dimension for recommendation, combining richer information about
historical encounters with the benefits of retaining more events.

\bibliographystyle{ACM-Reference-Format}
\bibliography{references}
\end{document}